\documentclass[lettersize,journal]{IEEEtran}

\usepackage{amsmath,amsfonts}
\usepackage{algorithmic}
\usepackage{algorithm}
\usepackage{array}
\usepackage{subfig}

\usepackage{tikz}
\usepackage{todonotes}
\usepackage[para,online,flushleft]{threeparttable}
\usepackage{multirow}
\usepackage{hyperref}
\usepackage[capitalise]{cleveref}

\def\crimebb{CrimeBB }

\begin{document}

\label{main:paper}

\title{A Graph-based Stratified Sampling Methodology for the Analysis of (Underground) Forums}

\author{Giorgio Di Tizio,
        Gilberto Atondo Siu,
		Alice Hutchings,
		Fabio Massacci
	
	\thanks{\IEEEcompsocthanksitem G. Di Tizio (corresponding author) is with University of Trento, Italy.\protect\\E-mail: giorgio.ditizio@unitn.it
	\IEEEcompsocthanksitem Gilberto Atondo Siu and Alice Hutchings are with University of Cambridge, United Kingdom.
	\IEEEcompsocthanksitem F. Massacci is with University of Trento, Italy and Vrije Universiteit Amsterdam, The Netherlands.}
 
		\thanks{The final version of this paper appears in: IEEE Transactions on Information Forensics and Security, 2023. DOI: 10.1109/TIFS.2023.3304424}
	}



\maketitle

\begin{abstract}
[Context] Researchers analyze underground forums to study abuse and cybercrime activities. Due to the size of the forums and the domain expertise required to identify criminal discussions, most approaches employ supervised machine learning techniques to automatically classify the posts of interest. [Goal] Human annotation is costly. How to select samples to annotate that account for the structure of the forum? [Method] We present a methodology to generate stratified samples based on information about the centrality properties of the population and evaluate classifier performance. [Result] We observe that by employing a sample obtained from a uniform distribution of the post degree centrality metric, we maintain the same level of precision but significantly increase the recall (+30\%) compared to a sample whose distribution is respecting the population stratification. We find that classifiers trained with similar samples disagree on the classification of criminal activities up to 33\% of the time when deployed on the entire forum.
\end{abstract}


\section{Introduction}\label{sec:introduction}
\IEEEPARstart{U}{nderground} forums contain valuable information related to cybercriminal activities. Vast collections of underground forums provide insights into the daily activities of millions of users~\cite{branwen_dark_2015,azsecure-dataorg_azsecure_nodate,DBLP:conf/www/PastranaTHC18}. These datasets contain collections going back over 20 years, in multiple languages, with hundreds of millions of posts. The potential benefits to academics across multiple disciplines to address societal challenges are enormous.

However, the sheer volume  of discussions in threads and posts poses a challenge to researchers~\cite{hughes2021too}. Current approaches rely on keyword searches and machine learning (ML) algorithms to identify and classify discussions. Increasingly, studies~\cite{akyazi2021measuring,caines2018automatically,DBLP:conf/www/PortnoffADKBMLP17,DBLP:conf/uss/WegbergTSAGKCE18,zhou2022automated} are using supervised ML algorithms due to the increased accuracy where classifiers are trained on human-labeled data. 

While off-the-shelf Natural Language Processing (NLP) tools struggle with the domain specificity required~\cite{caines2018automatically}, human-labeling of data is a resource-intensive process, particularly as multiple annotators are required. For cybercrime forums, where jargon and specialised language abounds, annotators also require domain expertise. Therefore, there is a need for identifying the best use of limited resources. The choice of the sample data is a key feature that can impact the performance of the ML classifier. Current approaches randomly pick posts to annotate on a subset of the forum that is promising for the topic to investigate. 

One challenge for researchers when sampling in relation to illicit activities is that classes are generally imbalanced~\cite{DBLP:conf/eurosp/SiuCH21}. This is partly due to the vast majority of users being  active for only transient periods of time~\cite{hughes_digital}. Furthermore, there is a  vast range of discussion topics -- licit as well as illicit. This means researchers seeking to create tools to research the activities of more sustained users or less commonly discussed topics require large annotation samples, and sometimes creative sampling methods.

In this paper, we investigate how the performance of the classifier is impacted by sampling methodology. In particular, we propose a methodology to generate stratified samples based on network centrality metrics and compare the classifier performances. 

We address the following research questions:
\begin{itemize}
	\item \textbf{RQ1:} What are the changes in performance for a ML classifier using different centrality metrics to generate stratified training samples? 
	\item \textbf{RQ2:} What are the changes in performance using a different proportion compared to the population for the stratified training sample?
\end{itemize}

The paper makes the following contributions:
\begin{itemize}
	\item A graph database (DB) representing the structure and interactions in an underground forum. We release the anonymized structure on Zenodo~\cite{giorgio_di_tizio_2023_8228689} to facilitate data analysis and future research. Due to ethical reasons, the access to the actual content stored in the graph (posts, thread, and member names) is subject to a formal data sharing agreement with the Cambridge Cybercrime Centre.\footnote{https://www.cambridgecybercrime.uk/process.html}
	\item A methodology for the generation of stratified samples based on graph metrics to train ML classifiers and for the validation of their performance on the population.
	\item An analysis of the impact on ML classifiers performance due to changes in the samples characteristics.
\end{itemize}

\emph{Non-goals:} We are not interested in tuning the classifiers to obtain the best performance on a given sample. We do not aim to determine pitfalls in the design and implementation of experiments using ML systems~\cite{DBLP:conf/uss/PendleburyPJKC19,DBLP:journals/corr/abs-2010-09470}.

\section{Related Works}\label{sec:relatedworks}

\subsection{Analysis of Underground Forums}
Several works analyze underground forums to study specific areas of cybercrime. Common approaches rely on NLP and supervised ML using random samples for training and testing.

In the context of the classification of posts, Portnoff et al.~\cite{DBLP:conf/www/PortnoffADKBMLP17} proposed an automated classification of post type (buy, sell, and exchange currency), product offered/requested, and price. They also evaluated the performance of the supervised classifier by training and testing over eight forums. They observed that the tool performance significantly drops if used across forums. Similarly, Caines et al.~\cite{caines2018automatically} evaluated the performance of different statistical models and heuristics on labeling post type, author intent, and addressee in \texttt{Hack Forums (HF)}. Van Wegberg et al.~\cite{DBLP:conf/uss/WegbergTSAGKCE18} trained a support vector machine (SVM) classifier to identify the type of listings (e.g. cash-out, malware, remote access tools (RATs), accounts, etc.) discussed in eight marketplaces and determine their associated revenue. Atondo Siu et al.~\cite{DBLP:conf/eurosp/SiuCH21} trained a supervised ML systems to classify posts in \texttt{HF} into a class of crime (\emph{Non-criminal}, \emph{Access to system}, \emph{Bots \& Malware}, \emph{eWhoring}, \emph{Currency Exchange}, \emph{DDoS}, \emph{Identify Theft}, \emph{Spam}, \emph{Trading credentials}, \emph{VPN}) and analyzed the digital currency utilized in each class. They observed that there was a massive shift to Bitcoin after Liberty Reserve was taken down, and there was a demand for exchanging PayPal.

In the context of cybercrime-as-a-service (CaaS), Akyazi et al.~\cite{akyazi2021measuring} measured the typology of CaaS services in \texttt{HF} via a supervised ML classifier and observed only a few CaaS categories discussed extensively (botnet, reputation escalation, and traffic-as-a-service). Sun et al.~\cite{DBLP:conf/uss/SunOZRB00SDA21} investigated Concession-Abuse-as-a-Service, analysing the techniques employed in four underground forums. Bhalerao et al.~\cite{DBLP:conf/ecrime/BhaleraoASAM19} analyzed business-to-business interactions in two underground forums (\texttt{HF} and \texttt{Antichat}). They trained supervised ML classifiers to determine the product offered and the reply class (buying, selling, or other). From this classification, they built an interaction graph between members to determine the presence of supply chains in the criminal markets.

Several studies focus on predicting criminal activities based on forum discussions. Pastrana et al.~\cite{DBLP:conf/raid/PastranaHCB18} developed a SVM to classify posts of key actors in \texttt{HF} and predict which actors might be of interest to law enforcement. Van Wegberg et al.~\cite{van2020go} investigated vendors and product characteristics to predict products success via regression analysis. They observed a positive correlation with features like presence of refund policy, customer support, and use of vendor names.  Sun et al.~\cite{10.1145/3292006.3300036} studied the differences between private and public messages in underground forums and developed ML classifiers to predict presence of private interactions from public features. Yuan et al.~\cite{DBLP:conf/uss/YuanLL018} developed a tool to automatically identify new dark jargons in underground forum posts using NLP.

The availability of forum discussions over several years allows researchers to investigate the evolution of these ecosystems. Soska and Christin~\cite{soska-marketplace} performed a longitudinal analysis of 16 online marketplaces for two years to estimate the sales volume and the type of products exchanged. They observed significant gross income for big marketplaces like Silk Road in the order of hundreds of thousands of dollars per day. Furthermore, they observed that the volume of sales was not significantly impacted by the law enforcement take-downs.
Pastrana et al.~\cite{DBLP:conf/imc/PastranaHTT19} investigated eWhoring activities on underground forums for more than ten years. They trained a ML classifier to identify threads offering `packs' and then determine the origins of the images, the main actors actively engaged in this activity, and their profits. They built a social network graph of members active in eWhoring discussions and identified the key actors using the h-index and eigenvector centrality.
They observed how the interest of these users moves from gaming and hacking to market-related topics after the interation in eWhoring threads.
Allodi~\cite{DBLP:conf/ccs/Allodi17} investigated exploits traded in a Russian black market and their likelihood of exploitation in the wild. Similarly, Campobasso and Allodi~\cite{campobasso2023know} measured the market trade volume of user-profiles and attackers' purchasing preferences in an underground forum. Vu et al.~\cite{DBLP:conf/imc/VuHPCCSH20} performed a longitudinal analysis of the trading activities in \texttt{HF} and their evolution over different `eras'. They observed how currency exchange and payments account for the majority of the contracts and payments are performed using Bitcoin and PayPal.

Table~\ref{tab:sampling_techniques} summarize the papers by the sampling techniques employed to train classifiers.
\noindent\fbox{\parbox{0.48\textwidth}{\noindent
\emph{Current approaches for supervised ML training rely on a sample obtained through random sampling from interesting discussions and do not employ information of the "population" of the social network.}
}}
We instead propose a methodology to generate stratified samples employing centrality metrics from the forum structure.
\begin{table}\footnotesize
	\caption{State of the Art (SotA) on Sampling Technique for supervised classification of criminal activities in forums}
		\centering
		  \begin{threeparttable}
		\begin{tabular}{p{0.6\columnwidth} p{0.3\columnwidth}}
			\hline
			Sampling Technique &  Papers \\\hline
			Simple Random & \cite{DBLP:conf/uss/WegbergTSAGKCE18},\cite{DBLP:conf/raid/PastranaHCB18}, \cite{caines2018automatically}, \cite{10.1145/3292006.3300036}, \cite{DBLP:conf/ecrime/BhaleraoASAM19}, \cite{DBLP:conf/imc/PastranaHTT19}, \cite{DBLP:conf/eurosp/SiuCH21}, \cite{akyazi2021measuring},\cite{DBLP:conf/uss/SunOZRB00SDA21}\tnote{*} \\
		    Unspecified & \cite{DBLP:conf/www/PortnoffADKBMLP17}, \cite{soska-marketplace}\\
		    Stratified using network centrality metrics & Our work \\
			\hline
		\end{tabular}
		\begin{tablenotes}
        \item[*] Stratified Random sampling
        \end{tablenotes}
		\end{threeparttable}
		\label{tab:sampling_techniques}
\end{table}

\subsection{Social Network Analysis of Underground Forums}
Several works investigate the properties of online social networks to identify influential actors and analyze topics of interests~\cite{DBLP:conf/esorics/ZhaoAHM12,DBLP:conf/raid/PastranaHCB18,DBLP:conf/imc/PastranaHTT19,DBLP:conf/ecrime/HughesCH19,DBLP:conf/imc/VuHPCCSH20}. Motoyama et al.~\cite{motoyama2011analysis} analyzed social network dynamics from leaks of 6 underground forums. In particular, they generate social network relationships based on `friend' requests, private messages, and thread discussions. They evaluated how the social degree of these relations impacts their trading. Garg et al.~\cite{DBLP:conf/fc/GargAOG15} employed \emph{social network analysis} (SNA) to determine sub-communities inside forums, the correlation between centrality metrics for members, and the impact on the network structure of banned members. Pastrana et al.~\cite{DBLP:conf/raid/PastranaHCB18} employed SNA, logistic regression, and clustering to identify members that interact with known criminals and predict their likelihood of being involved in future criminal activities. Almukaynizi et al.~\cite{almukaynizi2017predicting} employed social network metrics as features to predict the exploitation of vulnerabilities discussed in forums.

SNA is extensively used to identify key actors in underground forums~\cite{DBLP:conf/raid/PastranaHCB18,DBLP:conf/icbk/ZhangFY0WXS18,DBLP:journals/ijdsn/HuangGGL21}.
Our centrality metric approach is similar to Pete et al.~\cite{DBLP:conf/eurosp/PeteHCB20}, who constructed undirected graphs of six underground forums based on 6 months of observation, computed network statistics, and analyzed the network structure. Using centrality metrics they identified important members in the network and performed a qualitative analysis of the topics covered. In contrast to our work, their focus is on insights into the structures of different small forum snapshots. We instead propose a methodology to identify relevant samples for training ML classifiers based on the characteristics of the population of the entire forum, which can span several years of observation. Analysis of social network is also extended to a multilayer network to capture the importance of members over different communication mediums. For example, Ficara et al.~\cite{fi14050123} created a network with three layers named `Meetings', `Phone Calls', and `Crimes' to describe interactions of Sicilian Mafia members and identify key actors over different layers.

\noindent\fbox{\parbox{0.48\textwidth}{\noindent
\emph{Social Network Analysis in underground forums is mainly focused on identifying key actors and field experts and how the social relations influence the criminal activities in these forums. Prior works ignored social network characteristics, like centrality, to select representative posts from the population.}}
}

\subsection{Sample Representativeness in Online Social Network}
Prior work has analyzed sampling techniques for online social networks to recover the properties of the network in case the entire social network cannot be used and a sample must be extracted (e.g. via Twitter API~\cite{morstatter2013sample}) or to extract representative samples or samples with specific characteristics, e.g. high degree centrality~\cite{DBLP:conf/pakdd/MaiyaB10}. These sampling techniques rely on random node extraction, random edges extraction, exploration via random walks~\cite{DBLP:journals/datamine/ZhaoWLTG19,DBLP:journals/jsac/GjokaBKM11}, and snowball sampling~\cite{goodman1961snowball}. Studies investigated how  the different sampling techniques conserve the ranking of nodes, the visibility of groups~\cite{DBLP:conf/www/WagnerSKPS17}, how robust different centrality metrics are~\cite{costenbader2003stability} and proposed variants to preserve the properties of the original network, in particular ratio of nodes and edges and topology, based on hierarchical community and densification power law~\cite{DBLP:journals/isci/YoonKHKP15}.
\noindent\fbox{\parbox{0.48\textwidth}{\noindent
\emph{This line of research focuses on extracting a representative subgraph from a large network by trying to maintain the unknown properties of the population.}
}}
We instead focus on extracting sample elements exploiting known network characteristics of the entire population to improve ML classification.

\section{Ethical Considerations}

This work requires individuals to read posts on a forum. We note the dataset is collected from the public Internet, and is used for research on collective behaviour, without aiming to identify particular members. Given that users in underground forums hide behind a username, it is not possible to obtain consent from users as that would require us to identify them first. In accordance with the Menlo Report~\cite{dittrich2012menlo}, we informed the ethics committee so that we could waive the requirement for informed consent. The ethics committee at the Department of Computer Science \& Technology, University of Cambridge,
considered and approved this research. We paraphrase our example posts to limit the likelihood that they will be attributed back to their original authors.

\section{Methodology}\label{sec:methodology}
\begin{figure*}
\captionsetup[subfigure]{labelformat=empty}
\subfloat[Our methodology consists of: \emph{1) Graph DB generation:} the forums (F), boards (B), threads (T), posts, and members (M) are represented in terms of nodes and edges to highlight relationships. \emph{2) Population extraction:} a subgraph of the forum(s) is identified for the generation of samples for the topic. The resulting subgraph, projected based on some rules, is the population from which the sample is generated. \emph{3) Distribution extraction:} A graph metric is applied to the sub-graph to determine the distribution of the feature on the population. \emph{4) Sample generation:} A sample that respects the stratification is created. In contrast, the current state of the art approaches filters the forum for interesting discussions and randomly extracts posts from the identified subset.]{%
\includegraphics[width=\textwidth]{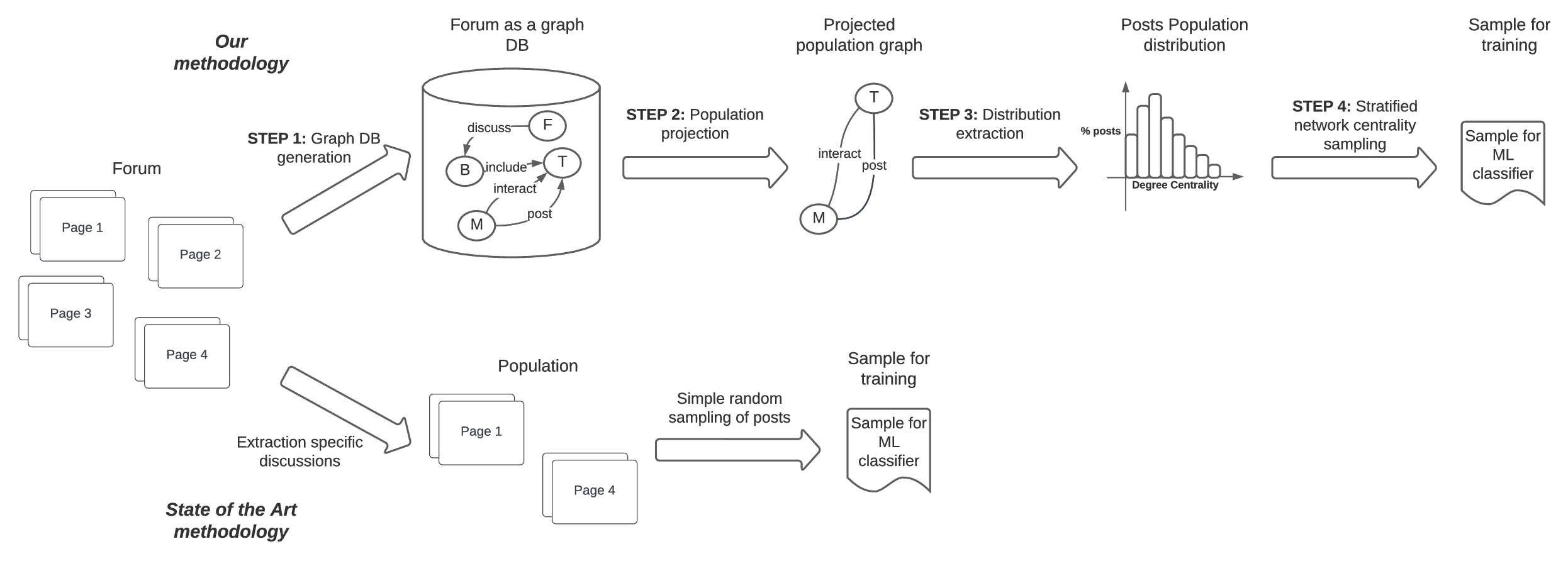}%
}
\centering
\caption{Methodology for stratified sampling of Forums}
\label{fig:methodology}
\end{figure*}

We present our methodology to generate stratified samples based on centrality metrics from a population of members in a forum. Fig.~\ref{fig:methodology} summarizes our overall procedure in comparison with the current state of the art approach.

\subsection{STEP 1: Graph Database generation}
We map a forum into a graph $G(V,E)$, in which $v \in V$ can be a board (B), a thread (T), or a member (M) node type and $e \in E$ can be one of the following relationships:
\begin{itemize}
    \item Forum$\xrightarrow{discuss}$Board : a forum discusses one or more general topics defined in boards. Boards can cover a broad range of topics related to Online Gaming, Cryptography, Reverse Engineering, RATs, etc.
    \item Board$\xrightarrow{include}$Thread: a board includes one or more member-contributed topics related to the general topic of the board.
    \item Member$\xrightarrow{post\{content, post\_type\}}$Thread: a member posts one or more comments inside a given thread. The comment is defined in the relationship property \emph{content}. The post is also classified for its intent, for example, an offer, a request for services, an exchange, or a tutorial.
    \item Member$\xrightarrow{interact\{weight\}}$Thread: a member interacts with one or more threads with a certain frequency given by the number of posts as described by the property \emph{weight}.
\end{itemize}
The schema of the graph DB is reported after the application of STEP 1 in~\cref{fig:methodology}. Nodes of the same type are connected through a different node type. For example, two members are connected if they post on the same thread or if they post in two different threads included in the same board.
The \emph{interact} edge is needed only for the graph analysis part. For performance reasons, this type of edge is created when the entire DB is generated.
Be $A$ the adjacent matrix of \emph{interact} and $W$ the matrix of weights associated with each \emph{interact} relationships. For simplicity, in the sequel we will use directly $G(V,E)$, $A$, and  $W$ as referring to the selected subset from the entire forum.

\subsection{STEP 2: Population projection}
Given a topic of interest for the analysis of cybercrime activities on a forum (e.g. eWhoring~\cite{DBLP:conf/imc/PastranaHTT19}, marketplace offered goods~\cite{DBLP:conf/uss/WegbergTSAGKCE18,DBLP:conf/eurosp/SiuCH21}, etc.) a ML classifier must be trained on a sample that contains, among the others, examples of the topic of interest.
Given that most activities in the underground forums are legitimate~\cite{DBLP:conf/raid/PastranaHCB18} and only a subset deals with criminal activities, the sample is typically obtained from a sub-graph of the entire forum that is promising and representative for the study of the topic. This sub-graph represents the \emph{population} from which a sample is generated for training and testing. The population is obtained following a selection rule, for example by identifying specific boards and threads that deal with the topic of interest via keyword searches.

In terms of graph representation, this approach produces a subset of the graph DB based on the selection rule.

\subsection{STEP 3: Distribution extraction}
We compute for each member of the population a value describing its posting activity based on a centrality metric. We then compute the distribution of posts inducted by the member's value on the metric. The post is the unit of the \emph{population} of interest for the analysis of cybercrime activities and composes the training sample for the ML model.

\subsubsection{Centrality Metrics over Members}

All metrics are computed for all nodes $v_{i} \in V$ where $v_{i}$ belongs to the \emph{Member} node type.

\paragraph{Post degree centrality} measures the amount of activity in terms of number of posts of the members composing the \emph{population}.
\begin{equation}
\mathit{post\_centrality(v_i)} = \sum_{j} W_{ij}*A_{ij}
\end{equation}

\paragraph{Thread degree centrality} measures the amount of distinct threads in which the population of members interact with. This metric differs from the post-degree centrality because it is used to discern members that are mainly active in few threads from members that interact in different discussions.

\begin{equation}
\mathit{thread\_centrality(v_i)} = \sum_{j} A_{ij}
\end{equation}

\paragraph{Eigenvector centrality} measures how much members participate in "hot" (highly participated) threads by looking at the importance of the thread nodes to which a member node is linked.
\begin{equation}
\resizebox{\columnwidth}{!}{$
\mathit{eigenvector\_centrality(v_i)} = \frac{1}{\lambda_{max}}\sum_{j} A_{ij}*\mathit{eigenvector\_centrality(v_j)}
$}
\end{equation}

\subsubsection{Distribution induced by centrality metrics}
Once the distribution of the centrality metric over the members is obtained, we compute the distribution of posts in the population \emph{induced} by the metric on the members and we normalize the distribution. The resulting distribution describes the percentage of posts associated with members of the population with a certain value for the centrality metric.
For all metrics, we observed 
few members with extreme values for a metric. For example, few members have a post-degree centrality greater than 10\,000.

The skewed distribution induced by the centrality metric can affect the sampling mechanisms. The size of the bins of the distribution must be adjusted to avoid biased sampling due to sample size. Suppose one wants to generate a sample of \emph{S} posts, the percentage of posts in each bin must be greater than $\frac{1}{S}$ such that at least one post can be picked from each bin. One would want to be able to pick at least 25 elements from each bin to have statistical significant results~\cite{agresti2016statistics}. To achieve this, we transform the distribution using the logarithm in base 10 and merge together into the same bin posts to have a percentage of the overall sample size greater or equal than $\frac{25}{S}$ posts. 

\subsection{STEP 4: Stratified Sample generation}
Based on the distribution of posts inducted by a centrality metric on the members, we generate stratified samples whose distribution respect the characteristics of the population. For example, suppose that the distribution of the posts of the population based on the \emph{post degree} centrality is such that 70\% of the posts are obtained from members with post degree centrality less than 10, 20\% of the posts are obtained from members with post degree centrality less than 100, and 10\% with less than 1\,000.
In particular, we generated 2 types of samples:
\begin{itemize}
    \item \emph{Proportional Sample}: that presents the same distribution (proportion) of the centrality metric as the population. The sample will be composed for 70\% of its size of posts from members that posted less than 10 posts, for 20\% of posts from members that posted less than 100, and for 10\% of posts from members that posted less than 1\,000.
    \item \emph{Uniform Sample}: that presents an uniform distribution of the centrality metric. Along the previous example, the new sample will present an equal number of posts from members that posted less than 10, less than 100, and less than 1\,000.
\end{itemize}

\begin{table*}\footnotesize
	\caption{Guideline Annotation Posts}
		\centering
		\begin{tabular}{p{0.25\columnwidth} p{0.85\columnwidth} p{0.8\columnwidth}}
			\hline
			Crime Type &  Description & Anonymized Example \\ \hline
		    Not criminal & Unrelated to crime. \emph{Including} sharing, selling of games points, skins, etc. & "Xbox One. Comment if you want to play.", "This post is to warm that the user X account was compromised by an hacker that also threat me"\\
		    Access to system & Exploitation of vulnerabilities (e.g. SQLi) where there is no innocent usage (e.g. pen-testing). \emph{Excluding} the use of malware. & "How to access a phone's text messages and calls without physical access to it." \\
		    Bots \& Malware & Botnet, malware, and related services. \emph{Excluding} social network bots. & "How to make my server file (of RAT) FUD????"\\
		    DDoS \& booting & DDoS attack and stress testing. \emph{Excluding} posts selling hosting with DoS protection. & "Would you be interested in investing in a SST service 100\% money would be made back plus more." \\
		    Spam & Spam, email sharing, or marketing services. The technique employed must be clearly stated (e.g. use of Adfly). \emph{Including} traffic generated, social network bots, request for views and subscribers.  & "Earn passive money with clickbank"\\
		    Trading credentials & Trading accounts including gaming and social network. \emph{Including} free accounts/credentials. \emph{Excluding} sell of domains, accounts in which the seller is the owner of the domain or service the accounts belong to. & "Selling sickest kik" \\
		    VPN \& hosting & VPN and hosting services. \emph{Including} requests and offers of VPN. & "I am looking for someone to host OMCPool.net in return for a share in the profits." \\
			\hline
		\end{tabular}
		\label{tab:guidelines_annotation}
\end{table*}

The generation of the sample can be subject to more constraints. For example, a maximum number of posts to include in the sample or the need to include available annotated posts (belonging to the population) to reduce the manual effort of the annotation.
In terms of annotation, a coding scheme and standardized procedure must be developed to label each sample. \cref{tab:guidelines_annotation} summarizes an example of the coding scheme for Atondo Siu et al.'s~\cite{DBLP:conf/eurosp/SiuCH21} crime type classifier, which provides anonymized examples for each class. If some classes are rare and the sample does not include enough posts for them, we ignore these rare classes and label them as part of the main class. Each sample must be independently annotated by at least two annotators to avoid subjective interpretation of the text. A metric like Cohen's or Fleiss's $\kappa$ must be employed to measure inter-annotator agreement. The posts with different annotations are reevaluated by all annotators jointly. The sample will be used to train the ML classifier.

\subsection{Validation of ML classifiers performance}
Once a ML classifier is trained on a sample, we evaluate its performance using an independent test sample that belongs to the same population.

To compare different sampling strategies, we directly run the classifiers on the entire population to determine the percentage of posts belonging to a class. We extract a random sample from the set of posts in which the two classifiers disagree respecting the stratification, annotate it, and evaluate the performance on that sample only. We then compute the Agresti Coull confidence interval (CI)~\cite{agresti1998approximate} to determine the range of agreement between classifiers on each class. The Agresti Coull CI is used when the sample size is greater than 40~\cite{10.1214/ss/1009213286}. Differences in performance between the two classifiers are only due to the set of posts in which the classifiers disagree.  

\section{Forum Dataset}
We relied on the \crimebb dataset~\cite{DBLP:conf/www/PastranaTHC18}, a database of underground forums available upon request.
We focused on \texttt{Hack Forums}, which is the largest and long-lived underground English-language forum, famous for the release of the Mirai botnet source code.
We represented the \texttt{HF} database in a graph DB using Neo4j\footnote{https://neo4j.com/}, a graph database that explicitly represents relationships between entities. The forum contains $\approx$680k members, with $\approx$42M posts over more than 4M threads.

We observed a user, likely the forum administrator, that presented an extreme number of posts in different threads. We observed that the posts were related to managing the forum and did not add any value to the analysis. We thus removed this member from the analysis. We further considered all threads and posts up to June 2018 to compare with a sample obtained from the related works~\cite{DBLP:conf/eurosp/SiuCH21}.

\section{Analysis}\label{sec:analysis}
We computed the centrality metrics using the Neo4j Graph Data Science (GDS) Library. The GDS library exploits an in-memory graph projected from the DB to efficiently run graph algorithms on large graphs.

We evaluate the performance of the classifiers over samples obtained using our methodology on different centrality metrics and compare the performance with a sample obtained using random sampling from Atondo Siu et al.~\cite{DBLP:conf/eurosp/SiuCH21} from the same period of time. To reduce the manual effort of the annotation, we constrained the generation of the new samples by keeping as many entries from the random sample as possible that respect the distribution of the centrality metric considered.

We first identified the population from which the random sampling of posts from~\cite{DBLP:conf/eurosp/SiuCH21} has been extracted to compute the population centrality metrics. The sample in~\cite{DBLP:conf/eurosp/SiuCH21} is composed of several samples:
\begin{itemize}
    \item \textbf{General HF Random Sample:} 500 posts extracted randomly from the entire \texttt{HF}.
    \item \textbf{Trading  HF Random Sample:} 1\,500 posts extracted randomly from all posts in \texttt{HF} classified as `Offer', `Request', `Exchange', and `Tutorial' by Caines et al.~\cite{caines2018automatically}\footnote{The  sample for training used in~\cite{caines2018automatically} is based on the following boards: \emph{Beginner Hacking}, \emph{Premium Sellers}, and additional 13 boards chosen at random (\emph{Computer and Online Gaming; Cryptography and Encryption Market; Decompiling, Reverse Engineering, Disassembly, and Debugging; Domain Trading; Ebook Bazaar; HF API; Marketplace Discussions; Remote Administration Tools; Secondary Sellers Market; Shopping Deals; Web Browsers; Windows 10; World of Warcraft})}.
    \item \textbf{Currency HF Random Sample:} 2\,000 posts extracted randomly from all posts in \texttt{HF} from members that published at least one post in the \emph{Currency Exchange} board.
\end{itemize}
We focus the analysis on the \emph{Trading HF Random Sample} because it can be easily described by the application of a selection rule and it is general enough to include discussions on different crime topics. We thus identified the population from which the sample was obtained:
\begin{itemize}
    \item \textbf{Trading HF Population:} The subgraph obtained by the threads in \texttt{HF} and their \emph{subset} of posts and members that are classified as `Offer', `Request', `Exchange', and `Tutorial' by Caines et al.~\cite{caines2018automatically}.
\end{itemize}

Tab.~\ref{tab:population_statistics} summarizes the selection rule for the block and the characteristics of the graph population. \cref{tab:samples_summary} summarizes the samples for the analysis and their sampling strategy.

\begin{table}\footnotesize
	\caption{Population graph Statistics}
		\centering
		\begin{tabular}{
		p{0.55\columnwidth} p{0.15\columnwidth} p{0.15\columnwidth}}
			\hline
			 Selection Rule &  \# Nodes &   \# Edges  \\\hline
			\emph{Subset} of posts and members that are classified as 'Offer','Request','Exchange', and 'Tutorial' by~\cite{caines2018automatically} & $\approx$447k (member), $\approx$3M (thread) & $\approx$11.3M (post), $\approx$9.6M (interact)  \\
			\hline
		\end{tabular}
		\label{tab:population_statistics}
\end{table}

\begin{table}\footnotesize
	\caption{Samples from Trading HF Population}
		\centering
		\begin{tabular}{
		p{0.35\columnwidth} p{0.41\columnwidth} r}
			\hline
			Name & Sampling Strategy & \# Posts \\\hline
            Trading HF random & Simple random sampling from population & 1\,500\\
            Post degree Proportional & Proportional stratified sampling from post degree population distribution & 1\,500\\
            Thread degree Proportional & Proportional stratified sampling from thread degree population distribution & 1\,500\\
            Eigenvector Proportional & Proportional stratified sampling from eigenvector population distribution & 1\,500\\
            Post degree Uniform & Uniform stratified sampling from post degree population distribution & 1\,500\\
			\hline
		\end{tabular}
		\label{tab:samples_summary}
\end{table}

\begin{figure*}
  \centering
  \subfloat[Distributions Post degree centrality]{
    \includegraphics[width=0.3\textwidth]{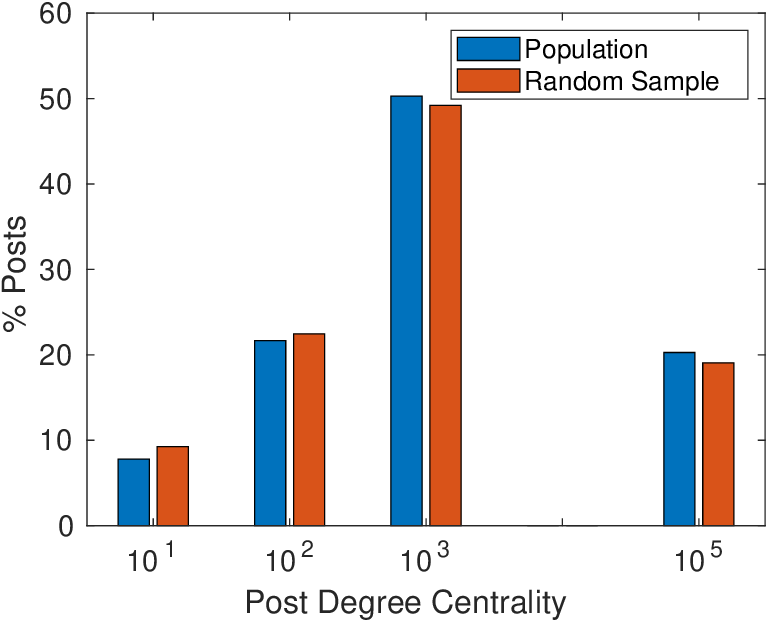}\label{fig:fig:distribution_posts}
    }
  \hfill
  \subfloat[Distributions Thread degree centrality]{
    \includegraphics[width=0.3\textwidth]{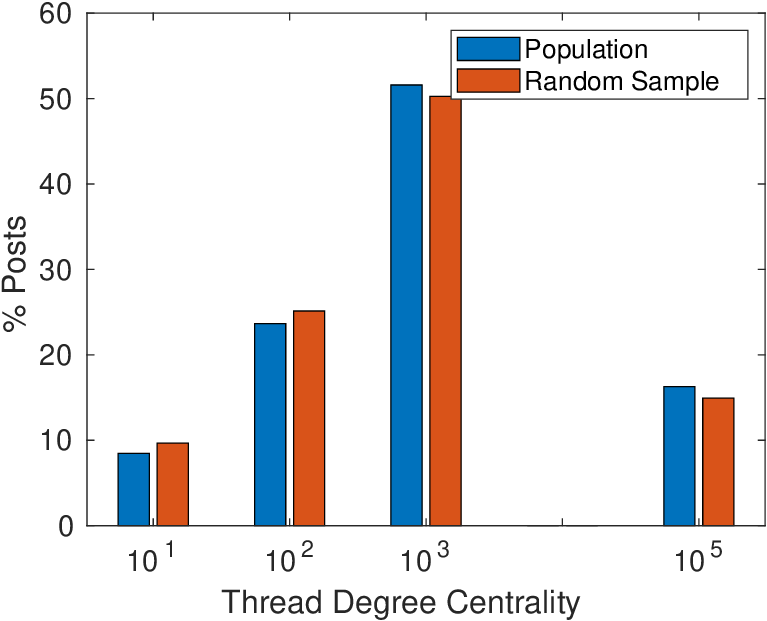}\label{fig:fig:distribution_threads}
    }
    \hfill
  \subfloat[Distributions Eigenvector centrality]{
    \includegraphics[width=0.3\textwidth]{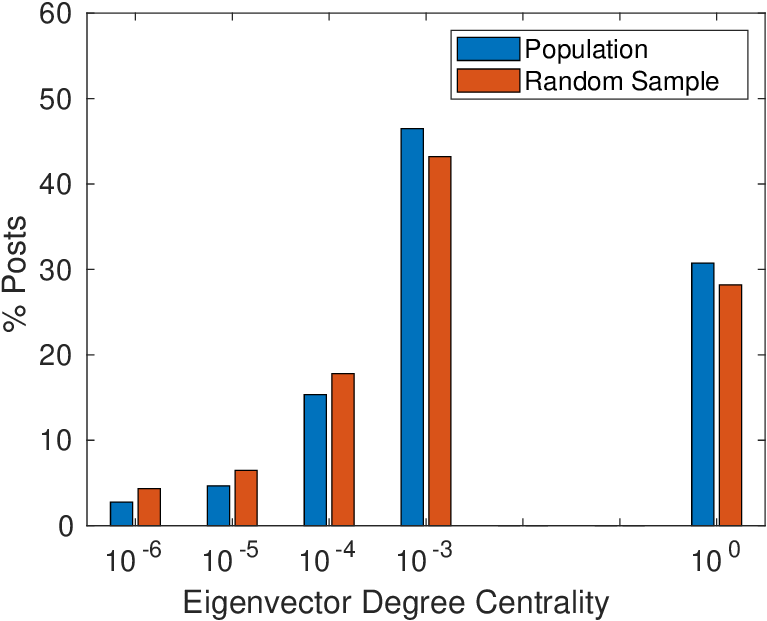}\label{fig:fig:distribution_eigenvector}
    }
    \caption{Distribution of population and Trading HF random sample over different centrality metrics. The distribution is obtained by merging together bins to obtain bins with enough posts to be sample given the sample size. All posts in a bin belongs to members whose centrality metric is strictly less than the value on the bin.}
    \label{fig:distribution_centrality_metrics}
\end{figure*}

\cref{fig:distribution_centrality_metrics} shows the distribution of the centrality metrics for the posts of the \emph{Trading HF} population and the 1\,500 posts of the Trading HF random sample from Atondo Siu et al.~\cite{DBLP:conf/eurosp/SiuCH21}. Each bin contains the posts associated with members whose centrality metric is strictly less than the bin value on the x-axis. The Trading HF random sample distribution is already similar to the population distribution because the sample size is large enough. For example, the probability \emph{p} that a post belongs to a member with less than 10 posts is $\approx$8\% (\cref{fig:fig:distribution_posts}) for the population. Thus, the standard error of the sample proportion is given by $\sqrt{\frac{p(1-p)}{n}}$=0.0137 for \emph{n}=1\,500. We expect the sample proportion to be within 0.92$\pm$0.0137 and that is the case for our Trading HF random sample that presents a $\hat{p}$=0.908.

\subsection{Annotation and Classes}
We obtained the 1\,500 posts sample of the Trading HF random sample from Atondo Siu et al.~\cite{DBLP:conf/eurosp/SiuCH21} and used the same coding scheme and procedure for the new samples. We employed the coding scheme in ~\cref{tab:guidelines_annotation}.  We manually classified the posts in one of the classes of crime as described in \cref{tab:class_types}.
Atondo Siu et al.~\cite{DBLP:conf/eurosp/SiuCH21} considered a larger set of criminal types. However, by looking at the classes in Atondo Siu's sample, we did not find enough instances for all classes\footnote{Indeed the analysis in~\cite{DBLP:conf/eurosp/SiuCH21} is performed with the addition of the 500 posts from the \emph{General HF} sample and the 2000 posts from the \emph{Currency HF} sample}. For a first approximation, we ignored these rare classes and classify them as \emph{not criminal}. This approximation does not affect the comparison of the performance between samples because all samples are annotated with the same annotation procedure.  For all stratified samples, we reused as many posts as possible from the Trading HF random sample to reduce the manual effort.

Three researchers independently annotated the posts for the post degree, thread degree, and eigenvector \emph{proportional samples}. Two of them were involved in the initial annotation of the Trading HF random sample. The new annotated posts for each sample were 34, 39, and 90 respectively. We report the Fleiss's $\kappa$ that measures the agreement among multiple annotators. The Fleiss's $\kappa$ ranges from 0.68 to 0.79. A value greater than 0.6 indicates a substantial agreement~\cite{landis1977measurement}. 
Two researchers further annotated 432 posts to generate the 
\emph{uniform samples} for the post degree centrality metric. We report the Cohen's $\kappa$ that measures the agreement between two annotators. The Cohen's $\kappa$ is 0.74 for the uniform post degree sample.

\begin{table*}\footnotesize
	\caption{Crime type classes and \# posts per sample}
		\centering
		\begin{tabular}{p{0.25\columnwidth}p{0.6\columnwidth}rrrrr}
			\hline
			Crime Type &  Description & \multicolumn{1}{p{0.15\columnwidth}}{Trading HF random} & \multicolumn{1}{p{0.15\columnwidth}}{Post degr. Propor.} & \multicolumn{1}{p{0.15\columnwidth}}{Thread degr. Propor.} & \multicolumn{1}{p{0.15\columnwidth}}{Eigenvector Proport.}& \multicolumn{1}{p{0.15\columnwidth}}{Post degr. Uniform} \\ \hline
		    Not criminal & Unrelated to crime & 1041 & 1048 & 1050 & 1055 & 983  \\
		    Access to system & Exploitation of vulnerabilities (e.g. SQLi) & 57 & 56 & 53 & 56 & 78\\
		    Bots \& Malware & Bots, malware, and related services & 157 & 156 & 154 & 144 & 164  \\
		    DDoS \& booting & DDoS attack and stress testing & 59 & 59 & 57 & 62 & 63\\
		    Spam & Spam, email sharing, or marketing services & 46 & 43 & 46 & 42 & 61 \\
		    Trading credentials & Trading accounts & 106 & 104 & 105 & 103 & 115 \\
		    VPN \& hosting & VPN and hosting services & 34 & 34 & 35 & 38 & 36 \\
			\hline
		\end{tabular}
		\label{tab:class_types}
\end{table*}

\subsection{Training ML models}
We evaluated the performance using the XGBoost and the SVM models. The SVM model is a simple model commonly used to classify posts in underground forums~\cite{DBLP:conf/uss/WegbergTSAGKCE18,DBLP:conf/raid/PastranaHCB18,DBLP:conf/www/PortnoffADKBMLP17,soska-marketplace}, while the XGBoost is
a state-of-the-art model that showed promising results in classifying posts in underground forum~\cite{caines2018automatically,DBLP:conf/eurosp/SiuCH21,akyazi2021measuring} and it is less subject to overfitting of training data compared to other classifiers~\cite{DBLP:conf/eurosp/SiuCH21}. We thus measure if the effect of different sampling is observable over different ML algorithms.
We consider as input to the classifier a text composed of the post content, the thread title, and the board title from the forum. We pre-processed the text to convert capitalized letters to lowercase, remove stop-words, tokenize the input, and lemmatize the words using the NLTK library~\cite{nltk}. We extract a vector of lexical features using \emph{tf-idf}. Given that most posts in the sample are classified as \emph{not criminal}, we re-sampled the training set using SMOTE~\cite{chawla2002smote} to overcome imbalances among classes.

\section{Performance}
We now report the ML classifier performance trained on the different samples and apply our methodology to validate their performance on a representative sample of the population. For the crime type classes \emph{DDoS} and \emph{Spam} we observed that the classifier did not predict any post in 53\% of the repeated stratified holdout for the Trading HF random sample due to the inability of the classifier to recognize samples of these classes. Similarly for the Post degree, Thread degree, and Eigenvector proportional samples, where classifier did not predict any post as \emph{DDoS} or \emph{Spam} in 36\%, 46\%, and 26\% of the stratified holdout runs respectively. This behavior is probably due to a lack of posts for these classes in the sample that allows the classifier to extract significant characteristics. Any attempt to retrospectively get more data on those classes will bias the sample by construction and thus, to avoid unfair performance comparison, we ignored \emph{DDoS} and \emph{Spam} in the analysis.

\subsubsection{Proportional Samples via Centrality Metrics}\label{sec:analysis:performance:shared_test}
We now address \textbf{RQ1} by looking at the classifier performance using the different centrality metrics to generate training samples.

We compare the performance of the ML models using as training set the Trading HF random sample from Atondo Siu et al.~\cite{DBLP:conf/eurosp/SiuCH21} and the post degree, thread degree, and eigenvector \emph{proportional} stratified sample. For testing we used a new 500 posts test sample obtained randomly from the population, annotated by two researchers (Cohen's $\kappa$=0.75).
We performed a repeated stratified holdout for the training and test sample with 30 different random seeds and average the results using the geometric mean. The geometric mean is best suited to summarize ratios~\cite{fleming1986not}. The stratification in the holdout depends on  the sample used for training. For example, when using the Trading HF random sample, the stratification is based on its class distribution\footnote{This is the common approach performed by the state-of-the-art.},  when using the post degree sample the stratification is  based on the population post-degree centrality distribution, and similarly for the other samples.

\cref{tab:xgboost_performance,tab:svm_performance} show the performance and the absolute and relative change considering the Trading HF random sample as the reference value in performance for the XGBoost and SVM models.
\noindent\fbox{\parbox{0.48\textwidth}{\noindent
\emph{We did not observe significant differences in the overall precision and recall using post degree, thread degree, and eigenvector centrality metrics for stratified sampling compared to the random sample.}
}}
This can be explained by the fact that the Trading HF random sample already presents a distribution similar to the population distribution for all centrality metrics (see \cref{fig:distribution_centrality_metrics}) and the change in the number of posts compared to the Trading HF random sample is small (2.2\% for the post degree, 2.6\% for the thread degree, and 6\% for the eigenvector \emph{proportional} stratified sample respectively). There are no significant differences between the two classifiers too, as the order of magnitude of precision and recall did not change.

\begin{table*}\footnotesize
     \centering
     \caption{Comparison XGBoost performance using Trading HF random and proportional stratified samples}
     \footnotesize
     \begin{tabular}{p{0.25\columnwidth} p{0.12\columnwidth}p{0.18\columnwidth}p{0.20\columnwidth}p{0.20\columnwidth}|p{0.12\columnwidth}p{0.15\columnwidth}p{0.20\columnwidth}p{0.20\columnwidth}}
     \hline 
    & \multicolumn{4}{c|}{Precision} & \multicolumn{4}{c}{Recall}\\
    Class & Random & Post degree & Thread degree & Eigenvector & Random & Post degree & Thread degree & Eigenvector  \\ \hline
     Not Criminal & 0.78 & 0.77 & 0.78 & 0.78 & 0.93 & 0.94 & 0.93 & 0.93 \\
     Access to system & 0.54  & 0.52  & 0.57 & 0.51 & 0.24  & 0.24 & 0.24 & 0.23 \\
     Bots \& Malware & 0.65 & 0.68 & 0.61 & 0.67 & 0.53 & 0.51 & 0.54 & 0.54 \\
     Trading credentials & 0.59 & 0.61 & 0.64 & 0.60 & 0.58 & 0.56 & 0.57 & 0.53\\
     VPN \& hosting & 0.54 & 0.54 & 0.54 & 0.54 & 0.23 & 0.23 & 0.23 & 0.27 \\ \hline
    Geometric Mean & 0.61  &  0.62 & 0.62 & 0.61  & 0.44 & 0.43 & 0.44 & 0.44\\
    Absolute Change & / &  +0.01 & +0.01 & 0.00 & / & -0.01 & 0.00 & 0.00\\
    Relative Change & / & +1.6\% & +1.6\% & 0\%  & / & -2.3\% & 0\% & 0\% \\ \hline
    \end{tabular}
     \label{tab:xgboost_performance}
\end{table*} 

\begin{table*}\footnotesize
     \centering
     \caption{Comparison SVM performance using Trading HF random and proportional stratified samples}
     \footnotesize
     \begin{tabular}{p{0.25\columnwidth} p{0.13\columnwidth}p{0.15\columnwidth}p{0.20\columnwidth}p{0.20\columnwidth}|p{0.13\columnwidth}p{0.15\columnwidth}p{0.20\columnwidth}p{0.20\columnwidth}}
     \hline 
    & \multicolumn{4}{c|}{Precision} & \multicolumn{4}{c}{Recall}\\
    Class & Random & Post degree & Thread degree & Eigenvector & Random & Post degree & Thread degree & Eigenvector  \\ \hline
     Not Criminal & 0.78  & 0.77  & 0.78 & 0.78 & 0.93 & 0.94 & 0.94 & 0.93\\
     Access to system & 0.49 & 0.51 & 0.54 & 0.47 & 0.31 & 0.31 & 0.30 & 0.29\\
     Bots \& Malware & 0.74 & 0.76 & 0.73 & 0.79 & 0.53 & 0.50 & 0.54 & 0.53\\
     Trading credentials & 0.62 & 0.66 & 0.64 & 0.63 & 0.49 & 0.47 & 0.49 & 0.45\\
     VPN \& hosting & 0.58 & 0.62 & 0.62 & 0.63 & 0.35 & 0.33 & 0.36 & 0.42 \\ \hline
    Geometric Mean & 0.63  & 0.66  & 0.66 & 0.65 & 0.48 & 0.47 & 0.49 & 0.49\\
    Absolute Change & / & +0.03 & +0.03 & +0.02 & / & -0.01 & +0.01 & +0.01 \\
    Relative Change & / & +4.8\% & +4.8\% & +3.2\% & / & -2.1\% & +2.1\% & +2.1\%\\ \hline
    \end{tabular}
     \label{tab:svm_performance}
\end{table*} 

\subsubsection{Proportional vs Uniform Sample}
To address \textbf{RQ2} and determine if the distribution of the centrality metric plays a significant role in the ML performance, we trained the XGBoost and SVM model with a sample that significantly differs in the distribution of a centrality metric compared to the population. We generated using our methodology (\S\ref{sec:methodology}) a \emph{uniform} sample based on the post degree centrality metric and compared it with the performance of the model trained with the same centrality metric but using the \emph{proportional} sample.

\Cref{tab:xgboost_uniform_performance,tab:svm_uniform_performance} show the results and the relative change compared to the proportional sample, whose results are reported in \cref{tab:xgboost_performance,tab:svm_performance} respectively for XGBoost and SVM. 
\noindent\fbox{\parbox{0.48\textwidth}{\noindent
\emph{
We observed that the overall precision using the uniform sample is the same as the proportional sample but in contrast the recall significantly improves on both models (+30\% for XGBoost and +21\% for SVM).
}}}
Thus the positive effect on the recall of a uniform stratification sample is observable using different ML algorithms. 
  \begin{table}\footnotesize
     \caption{XGBoost performance using post degree uniform sample and relative change compared to post degree proportional sample from Table~\ref{tab:xgboost_performance}}
     \centering
     \footnotesize
     \begin{tabular}{p{0.25\columnwidth} p{0.3\columnwidth}p{0.3\columnwidth}}
     \hline 
    Class & Precision & Recall\\ \hline
     Not Criminal & 0.83 & 0.89 \\
     Access to system & 0.52  &  0.40 \\
     Bots \& Malware & 0.64 &  0.66\\
     Trading credentials & 0.59 & 0.64 \\
     VPN \& hosting & 0.54 & 0.36 \\ \hline
    Geometric Mean & 0.62   & 0.56 \\
    Absolute Change & 0.00 & +0.13 \\ 
    Relative Change & +0\% & +30\% \\\hline
    \end{tabular}
     \label{tab:xgboost_uniform_performance}
\end{table} 

  \begin{table}\footnotesize
     \caption{SVM performance using post degree uniform sample and relative change compared to post degree proportional sample from Table~\ref{tab:svm_performance}}
     \centering
     \footnotesize
     \begin{tabular}{p{0.25\columnwidth} p{0.3\columnwidth}p{0.3\columnwidth}}
     \hline 
    Class & Precision & Recall\\ \hline
     Not Criminal & 0.83 & 0.88\\
     Access to system & 0.46 & 0.40\\
     Bots \& Malware & 0.67 & 0.63\\
     Trading credentials & 0.62 & 0.58 \\
     VPN \& hosting & 0.61 & 0.53 \\ \hline
    Geometric Mean & 0.63 & 0.58 \\
    Absolute Change & 0.00 & +0.11\\
    Relative Change & +0.00\% & +21\%\\\hline
    \end{tabular}
     \label{tab:svm_uniform_performance}
\end{table}

\subsubsection{Agreement between classifiers}
The differences in performance between the Trading HF random and the post degree proportional stratified sample are relatively small if one only looks at \cref{tab:xgboost_performance,tab:svm_performance}. However, this difference can be significant when the classifier is deployed to classify an entire forum with millions of posts. We investigated the real impact of this small variation by running the classifiers trained on the Trading HF random and post degree \emph{proportional} sample on all population posts (\cref{tab:population_statistics}). We then computed the proportion of posts that were classified the same by both classifiers and the Agresti Coull CI to determine the range of agreement.
\cref{tab:xgboost_disagreement_random_post,tab:svm_disagreement_random_post} show the per-class agreement and disagreement for the Trading HF random sample and post-degree \emph{proportional} sample. 
\noindent\fbox{\parbox{0.48\textwidth}{\noindent
\emph{
The agreement for the crime type classes ranges between 67\% and 77\% for XGBoost and between 81\% and 85\% for SVM, thus although trained with very similar samples, the classifiers trained with a post degree proportional stratified sample and a random sample differ up to 1 out of 3 posts for certain crime type classes when deployed on the entire forum.}
}}

\begin{table}\footnotesize
    \caption{XGBoost Agreement Trading HF Random and Post-degree proportional classifiers.}
     \centering
     \footnotesize
     \resizebox{\columnwidth}{!}{%
     \begin{tabular}{p{0.25\columnwidth}rrrc}
     \hline 
     Class &
     \#Agree & \#Random only & \#Proportional only & CI Agreement \\ \hline
     Not Criminal  & 9\,219\,317 & 213\,491 & 294\,270 & (0.95,0.95)\\
     Access to system & 120\,068 & 25\,986 & 32\,058 & (0.67,0.68)\\
     Bots \& Malware & 561\,925 & 103\,996 & 67\,620 & (0.76,0.77)\\
     Trading credentials & 715\,415 & 138\,050 & 80\,540 & (0.76,0.77)\\
     VPN \& hosting & 123\,767 & 22\,998 & 26\,924 & (0.71,0.71)\\ \hline
    \end{tabular}
    }
    \label{tab:xgboost_disagreement_random_post}
\end{table} 

\begin{table}\footnotesize
    \caption{SVM Agreement Trading HF Random and Post-degree proportional classifiers.}
     \centering
     \footnotesize
     \resizebox{\columnwidth}{!}{%
     \begin{tabular}{p{0.25\columnwidth}rrrc}
     \hline 
     Class &
     \#Agree & \#Random only & \#Proportional only & CI Agreement \\ \hline
     Not Criminal  & 9\,010\,428  & 146\,372 & 285\,124 & (0.95,0.95) \\
     Access to system & 209\,395 & 31\,843 & 17\,830 & (0.81,0.81)\\
     Bots \& Malware & 576\,132  & 81\,901 & 47\,948 & (0.81,0.82)\\
     Trading credentials & 734\,492  & 119\,406 & 72\,089 & (0.79,0.79)\\
     VPN \& hosting & 119\,358  & 14\,676 & 7\,279 & (0.84,0.85)\\ \hline
    \end{tabular}
    }
    \label{tab:svm_disagreement_random_post}
\end{table} 

To provide some qualitative understanding of the differences we investigated the performance on those posts in which the two XGBoost classifiers disagree. 
We randomly picked for each class, 100 posts for which the classifiers disagreed.
A total of 500 posts among the disagreement were manually re-annotated to create a new test set to measure the performance of the classifiers on the disagreement (Cohen's $\kappa$=0.80). This test sample allows  one to investigate the region in which the boundaries of the classifier change due to the sample characteristics. \cref{tab:xgboost_test_vs_disagreement_performance} summarizes the 
classifiers performance on the test sample extracted from the disagreement posts. 

Although the overall performance are similar, in contrast to what observed in~\cref{tab:xgboost_performance} we have more significant differences in the precision and recall for certain crime type classes. For example, the Trading HF random sample has better precision for \emph{VPN \& hosting}, while the post degree proportional sample has better precision and recall for the \emph{Trading credentials}. Given that the performance on the posts in which the classifiers agreed is the same, either both are right or both are wrong, we expect the corresponding classifiers  to perform better in these classes when dealing with the entire forum because in the regions where the classifiers differ they have better performance. For example, the classifier trained with the post degree proportional sample incorrectly classified as \emph{VPN \& hosting} the following paraphrased example: \textit{"I made lots of money by hosting Minecraft on a server I rent"}\footnote{The real example is longer and rougher and would identify the user, even if we do not report the name here.},
while the classifier trained with the Trading HF random sample correctly did not classify it as related to VPN or hosting. Conversely the classifier trained with the Trading HF random sample classified as \emph{Trading credentials} the following paraphrased example: \textit{"Why would you want to unverify a Paypal account?"}, while the classifier trained with the post degree proportional sample correctly did not classify it as related to trading credentials.
\begin{table}\footnotesize
    \caption{XGBoost Disagreement Sample Performance.}
    \centering
     \footnotesize
    \resizebox{\columnwidth}{!}{%
     \begin{tabular}{p{0.25\columnwidth} p{0.1\columnwidth} p{0.25\columnwidth}| p{0.1\columnwidth} p{0.25\columnwidth}}
     \hline 
     & \multicolumn{2}{c|}{Precision} & \multicolumn{2}{c}{Recall} \\
    Class & Random & Post degree Prop. & Random & Post degree Prop.\\ \hline
     Not Criminal & 0.36 & 0.39 & 0.53 & 0.57\\
     Access to system & 0.39 & 0.38 & 0.33 & 0.33 \\
     Bots \& Malware & 0.58 & 0.61 & 0.53 & 0.43 \\
     Trading credentials & 0.41 & 0.56 & 0.49 & 0.65 \\
     VPN \& hosting & 0.68 & 0.53 & 0.47 & 0.49 \\ \hline
    Geometric Mean & 0.47 & 0.48 & 0.46 & 0.48\\ 
    Relative Change & / & +2.13\% & / & +4.35\% \\ \hline
    \end{tabular}
    }
     \label{tab:xgboost_test_vs_disagreement_performance}
\end{table} 

\section{Discussions and Limitations}\label{sec:discussion}
The overall practical objective of this research is to improve the results of the classifiers when they need to be trained on
human-labeled data obtained after a resource-intensive process by multiple annotators. In this respect, we found that the overall approach move in the right direction.
We summarized here the approximate cost, in terms of time, of the different phases for the preparation of the classifier:
\begin{itemize}
\item centrality metric $G(V=470k,E=9.6M)$: $t\approx1h$
\item manual annotation 1500 posts: $t\approx6h$
\item training 1500 posts: $t\approx15s$
\end{itemize}
The cost of human resources is still the dominant factor in our approach. However, additional metrics can be used to better select the samples that require manual annotation.

\subsection{Alternative Graph Metrics}

Other graph centrality metrics are considered in the literature. For example, Bramoullé et al.~\cite{10.1257/aer.104.3.898} considered the determination of network equilibrium points based on the smallest eigenvector, while Pete et al.~\cite{DBLP:conf/eurosp/PeteHCB20} used betweenness and closeness centrality to identify structural patterns between forums. We considered such metrics in the preliminary phases of our research as they would seem to provide additional key insights into the graph structures. However, we found them unsuitable for the purposes of the quick but large-scale analysis of cybercrime fora with millions of posts using effective ML methods to support limited resources. These metrics suffer from severe computational limitations when applied to a forum as large as ours. Indeed, the algorithm for finding the points of network equilibria based on the smallest eigenvector in \cite{10.1257/aer.104.3.898} is exponential, while \cref{eq:betweenness} used to compute the betweenness centrality requires first computing the number of shortest paths between pairs of nodes $s$ and $t$ ($\sigma(s,t)$), similar considerations for metrics like closeness and eccentricity.

\begin{equation}
    C_{B}(v_i) = \sum_{s,t\in V} \frac{\sigma(s,t|v)}{\sigma(s,t)}\label{eq:betweenness}
\end{equation}
In ~\cite{DBLP:conf/eurosp/PeteHCB20} the betweenness and closeness centrality were computed on a small fragment of the entire forum, and nonetheless this took a significant amount of time for the computation. Attempts to run Neo4j optimized algorithms on the entire graphs to compute betweenness centrality were aborted due to memory exhaustion after 24 hours on a machine with Intel Xeon Gold 5220R @ 2.20GHz with 64Gb JVM heap space.
 
The reason behind such poor performance is that cybercrime fora are massive (millions of nodes) connected graphs without a bounded diameter (the length of the longest shortest path between all pairs of nodes). As one can go from a member to a post, from this post to its thread, from this thread to another post by a different member, and from this member to another post in a different thread, etc. the graph is essentially connected. If two members are both very active, the shortest path between them can be bounded, e.g. because they both post on a popular thread. However, if two members rarely post something, then the shortest path between them could be very long and require traversing a large portion of the graph. Fortunately for society but unfortunately for graph algorithms, most members of a cybercrime forum are not active and post very little \cite{allodi2015then,DBLP:conf/imc/VuHPCCSH20}.

Since an exact algorithm to calculate betweenness centrality requires to compute first the shortest path across all nodes its computational complexity is $t\approx(m\cdot n)$. In our case, this would require $t\approx 4.3 \cdot 10^{12}$ operations and even by exploiting parallelization, it would be prohibitively expensive. The best algorithm for parallel single source shortest paths for power-law graphs similar to ours only gives a speedup of 2.4x on a 32-thread CPU~\cite{chi2022accelerating}. Even with such efficient algorithms, we would require data processing in the order of Teraflops. In hindsight, this computational hurdle explained how our attempts failed even on high-end servers.

Approximate algorithms, such as Brandes and Pich~\cite{brandes2007centrality}, that uses $k$ pivots per node, require the graph to have a bounded diameter so that the error is bounded by $\epsilon\cdot diam(G)$. However, in our case, the diameter is not bounded. Therefore, either the error increases to the point of being uninformative (remember that $diam(G)$ can be of several orders of magnitudes) or the number of pivots per node must be expanded until it reaches $n$ and thus its complexity has the same order of magnitude of the exact algorithm.

The costs of computing the centrality metrics would become the dominant factor and it is unclear whether it would bring better benefits than providing more manual annotations.

\subsection{Alternative Approach to Classifiers}

In \S\ref{sec:analysis} we relied on a previously obtained classification of post types to identify the population of interest (\emph{Trading HF} population). The population is thus affected by the precision and recall of the classifier by Caines et al.~\cite{caines2018automatically}. However, this does not influence the overall results because the centrality metrics are computed by considering classified posts as the entire and only population. 

Time biases in which future posts are used to predict previous posts~\cite{DBLP:conf/uss/PendleburyPJKC19} is not an issue because the aim is to classify the entire set of posts available at a certain point in time.

The classifiers trained in this paper with samples are not tuned to obtain the best achievable precision and recall. Thus the results are not directly comparable with the related work that aim at proposing the best-fitted classifier on a given dataset. However, they provide an indication of how the corresponding performance
 will change in the presence of optimization with respect to the use of unstratified samples in the training set.

Deep learning algorithms showed excellent performance in different classification tasks and seem promising for the classification of criminal discussions in underground forums~\cite{DBLP:conf/bigdataconf/DeliuLF17}. However, these algorithms require a huge number of training samples that, in this context, are obtained after a resource-intensive process by multiple annotators.

\section{Conclusions}\label{sec:conclusion}
We presented a methodology to generate new samples exploiting information about the centrality properties of the population and to evaluate the performance of the classifiers. We observed a significant increase in recall using a uniform distribution of the post degree centrality metric to generate the training sample. Other centrality metrics like thread and eigenvector did not differ in the overall performance. We also observed that the agreement between classifiers trained with similar samples can significantly disagree in their classification when the classifiers are deployed on the entire underground forum.
Further analysis using other distributions and centrality metrics such as betweenness centrality can only be applied to smaller  
fora to be computationally feasible (or require significant parallel computing resources). A different avenue for future work could be instead to use other graph analysis techniques such as clustering.
Future works can investigate the application of our stratified sampling methodology in a multilayer network~\cite{fi14050123}, where layers describe different interactions between members. For example, in case both private and public messages are available one can determine different sampling choices based on private and public interactions.

\section*{Acknowledgments}

This work is supported by 
the European Research Council (ERC) under the European Union’s Horizon 2020 research and innovation programme grant n.830929 (CyberSec4Europe) and n.952647 (AssureMOSS) for GDT and FM and grant n. 949127 (iCrime) for GAS and AH.
	\subsection*{CRediT statements}
	\emph{Conceptualization:}	GDT, FM; 
	\emph{Methodology:} GDT, FM; 	
	\emph{Software:} GDT, GAS; 	
	\emph{Validation:} GDT, FM, AH;	
    \emph{Formal analysis:} GDT, FM;	
    \emph{Investigation:} GDT, FM, AH;	
    \emph{Resources:} GAS, AH;	
    \emph{Data Curation:} GDT, GAS, AH; 	
    \emph{Writing - Original Draft:} GDT, FM	
    \emph{Writing - Review \& Editing:} GDT, FM, AH	
    \emph{Visualization:} GDT	
    \emph{Supervision:} FM, AH	
    \emph{Project administration:} FM, AH 
    \emph{Funding acquisition:} FM, AH 
\bibliographystyle{IEEEtran}
\bibliography{short,references}

\clearpage \onecolumn
	\appendix
	
\section*{Examples of Disagreement}
We provide some paraphrased posts in which the Trading HF random and post degree stratified sample disagreed. ~\cref{tab:disagreement_examples} shows the paraphrased examples, the classification of each classifier, and the annotation.

\begin{table*}[h]\footnotesize
	\caption{Examples of Disagreement between trained models}
		\centering
		\begin{tabular}{p{0.3\columnwidth} p{0.08\columnwidth} p{0.08\columnwidth} p{0.08\columnwidth} p{0.1\columnwidth} p{0.1\columnwidth} p{0.1\columnwidth}}
			\hline
			Example & Post degree Centrality & Thread degree Centrality & Eigenvector Centrality &  Classification Random  & Classification Post Degree & Annotation \\ \hline
		    "Help Keylogger. If you crypted you risk to made it unstable. It must be FUD to not be detected" & $3.8*10^{2}$ & $3.5*10^{2}$  & $2*10^{-4}$ & Bots \& Malware & Not criminal & Bots \& Malware \\
		    "To have a botnet: take a mp3 file, infect it and upload on a hosting to download" & $7.0*10^1$ & $6.8*10^1$ & $3.3*10^{-4}$ & VPN \& hosting & Bots \& Malware & Bots \& Malware \\
		    "This post is to warm that the user X account was compromised by an hacker that also threat me" & $1.1*10^{3}$ & $9.4*10^{2}$  & $1.6*10^{-3}$ & Not criminal & Trading credentials & Not criminal \\
		    "Hello, I updated this program to create a relaxing game. This is the virus scan for the program. Enjoy" & $5.3*10^2$ & $3.8*10^2$ & $2.1*10^{-4}$ & Bots \& Malware & Not criminal & Not criminal \\
		    "Hacking using IP. Where is the scanner link? Do I need to find it on the website?" & 3 & 3 & $6.04*10^{-7}$ & Not criminal & Access to system & Access to system \\
		    "I have a social network md5 password. I need to decrypt it. Any idea?" & $1.5*10^1$ & $1.5*10^1$ & $1.8*10^{-4}$ & Access to system & Bots \& Malware & Access to system \\
		    "Hosting service - DDoS protection and VPS hosting with 24/7 support. We accept requests. You will enjoy it" & $1.7*10^{2}$ & $1.4*10^{2}$ & $1.2*10^{-3}$ & Not criminal & VPN \& hosting & VPN \& hosting \\
		    "I am using these Proxies so I share with you [IP addresses]" & $1.3*10^1$ & $1.2*10^1$ & $6.2*10^{-4}$ & VPN \& hosting & Not criminal & VPN \& hosting\\ 
		    "How many Netflix accounts can you sell for 5 cents?" & $1.3*10^{2}$  & $1.1*10^{2}$  & $1.8*10^{-4}$ & Trading credentials & Not Criminal & Trading credentials \\
		    "I have thousands of accounts in [SOCIAL NETWORK] that have many contributions as requested by you" & $1.5*10^{2}$ & $1.4*10^{2}$ & $2.3*10^{-5}$ & Not criminal & Trading credentials & Trading credentials \\
			\hline
		\end{tabular}
		\label{tab:disagreement_examples}
\end{table*}

\end{document}